%% file: main.tex
\documentclass[letterpaper,twocolumn,10pt]{article}
\usepackage{usenix}
\microtypesetup{spacing=false}

\usepackage{booktabs} 
\usepackage{multirow}
\usepackage{tabularx}
\usepackage{xcolor}
\usepackage{tikz}
\usepackage{amsmath}
\usepackage{rotating}
\usepackage{pdflscape}  
\usepackage{capt-of}    
\usepackage{comment}

\usepackage[ruled,vlined,linesnumbered]{algorithm2e}

\usepackage{graphicx}

\newcolumntype{L}{>{\raggedright\arraybackslash}X}
\newcolumntype{s}{>{\hsize=0.55\hsize\raggedright\arraybackslash}X}
\newcolumntype{M}{>{\hsize=1.45\hsize\raggedright\arraybackslash}X}

\newsavebox{\takeawaybox}
\newenvironment{takeaway}
  {%
    \begin{center}
    \setlength{\fboxsep}{6pt}%
    \setlength{\fboxrule}{1pt}
    \begin{lrbox}{\takeawaybox}%
    \begin{minipage}{0.95\columnwidth}%
  }
  {%
    \end{minipage}%
    \end{lrbox}%
    \fcolorbox{black}{yellow!15}{\usebox{\takeawaybox}}%
    \end{center}
  }

\begin{document}

\date{}

\title{\Large \bf SoK: You Find What You Seek \\ Rethinking Oracles, Guidance, and Input Generation in Hardware Fuzzing }

\author{
    {\rm G Abarajithan}\\
    UC San Diego
    \and
    {\rm Zhenghua Ma}\\
    UC San Diego
    \and
    {\rm Cristian Tirelli}\\
    UC San Diego
    \and
    {\rm Andres Meza}\\
    UC San Diego
    \and
    {\rm Francesco Restuccia}\\
    UC San Diego
    \and
    {\rm Cynthia Sturton}\\
    University of North Carolina \\at Chapel Hill
    \and
    {\rm Ryan Kastner}\\
    UC San Diego
} 

\maketitle

\newcommand\zhenghua[1]{\textcolor{green}{\textbf{#1} -Zhenghua}}
\newcommand\ryan[1]{\textcolor{cyan}{\textbf{#1} -Ryan}}
\newcommand\aba[1]{\textcolor{purple}{\textbf{#1} -Aba}}
\newcommand\cynthia[1]{\textcolor{red}{\textbf{#1} -Cynthia}}
\newcommand\cristian[1]{\textcolor{blue}{\textbf{#1} -Cristian}}
\newcommand\francesco[1]{\textcolor{brown}{\textbf{#1} -Francesco}}
\newcommand\andy[1]{\textcolor{olive}{\textbf{#1} -Andy}}

\input{abstract}
\input{intro}

\input{framework}
\input{background}

\input{oracle}
\input{guidance}
\input{input_gen}
\input{benchmarks}

\input{how_to_use}

\input{discussion}

\appendix
\section{Open Science}

This paper is a Systematization of Knowledge based on an analysis
of publicly available literature. It does not report new experiments
or empirical measurements requiring source code, datasets, binaries,
or other experimental artifacts. The works analyzed in this paper
are identified in the bibliography.

\bibliographystyle{plainurl}
\bibliography{ref}

\end{document}

%% file: abstract.tex
\begin{abstract}
Hardware fuzzing is an active area in security verification research, yet its industrial adoption remains in its early stages.
This SoK examines which lessons from software fuzzing carry over to hardware and where unique approaches are needed.
By analyzing 52 fuzzers across RTL/IP, CPU, NoC, and SoC designs, we introduce an analytical framework that frames verification as a bounded search. 
This search is defined by its objective, oracle, guidance, input generation, target abstraction, and budget. 
Consequently, a campaign only uncovers failures it can effectively reach, recognize, and prioritize before exhausting its resources.
We distinguish two roles for hardware fuzzing: (1) augmenting constrained-random verification (CRV) via feedback-guided coverage and (2) directed adversarial testing based on threat models and security specifications.
Through our framework, we identify what each campaign can observe and generate, providing a basis for assessing the evidence behind reported results.
Our analysis suggests that mainstream adoption of hardware fuzzing will require reusable interfaces, target-specific verification assets, reproducible evaluations, and transparent reporting of cost and user effort.
\end{abstract}

%% file: intro.tex
\section{Introduction}
\label{sec:introduction}

Fuzzing is an established technique in software testing.
Google's OSS-Fuzz has helped identify and fix over 13,000 vulnerabilities and 50,000 bugs across 1,000 open-source projects~\cite{oss_fuzz}.
This success has motivated attempts to adopt fuzzing for hardware verification, especially in academic research.
Hardware fuzzing promises to automate verification and expose corner cases that constrained-random verification (CRV) and ad-hoc directed adversarial testing may miss in security-critical designs.
Despite this academic interest, hardware fuzzing has yet to be widely adopted in industrial hardware verification~\cite{hwfuzzenv}.

Software fuzzing began with automated random-input testing of UNIX utilities in 1990~\cite{sw_miller_1990}.
Over three decades, it matured through advances in failure detection, execution feedback, structured input generation, harnesses, and continuous fuzzing infrastructure~\cite{sw_art_science}.
Modern software fuzzing combines these components into workflows that can be deployed and maintained across many targets~\cite{aflplusplus,oss_fuzz}.

Hardware fuzzing faces its own challenges as hardware executes in parallel and bugs rarely crash a register-transfer-level (RTL) simulation.
Useful inputs also depend on whether the target is a CPU, an intellectual property (IP) block, or a full system-on-chip (SoC).
Oracles, guidance, and input generation from software fuzzers therefore do not always transfer.
Hardware fuzzing should adopt design patterns proven sustainable in software while adapting them to hardware verification objectives and existing verification assets.

In dynamic hardware security verification, hardware fuzzing can serve two objectives: (1) improving CRV by using feedback to steer test generation toward uncovered behaviors in a functional verification plan, and (2) performing directed adversarial testing by searching for an execution that violates a property derived from a security threat model.
In either role, the campaign must align its checks, progress signal, and generated actions with the verification plan, threat model or security specification.

Analyzing 52 hardware fuzzers across RTL/IP, CPU, network-on-chip (NoC), and SoC targets, this SoK develops a framework to ask (1) where fuzzing complements established verification, (2) what each fuzzer can recognize, reach, and prioritize, and (3) when hardware fuzzers can be compared, reproduced, and deployed.
Our contributions include:

\begin{itemize}
    \item We develop an analytical framework that treats hardware fuzzing as bounded search.
    It separates the oracle that recognizes failures, the input generator that determines reachability, and the guidance that prioritizes executions within a finite budget.
    It identifies the subcomponents, design choices, and tradeoffs in this decomposition.

    \item We analyze and compare 52 hardware fuzzers across IP, CPU, NoC, and SoC abstraction levels, considering their failure oracles, structural, functional, and directed guidance, input representations, mutation and scheduling policies, and solver-assisted target-to-input generation.
    
    \item We examine hardware-fuzzing guidance and identify a general mismatch between commonly used feedback signals and the functional and security behaviors encoded by verification plans and threat models.

    \item We compare evaluation, reproducibility, and deployment requirements.
    Drawing on software fuzzing and recent hardware-fuzzer evaluations~\cite{encarsia}, we identify conditions for fair comparison and mainstream adoption, including consistent metrics, versioned designs under test (DUTs), documented bug corpora, shared verification infrastructure, component ablations, and open artifacts.
\end{itemize}

%% file: framework.tex
\section{The Framework: You Find What You Seek}
\label{sec:framework}

\begin{figure}
    \centering
    \includegraphics[width=1\linewidth]{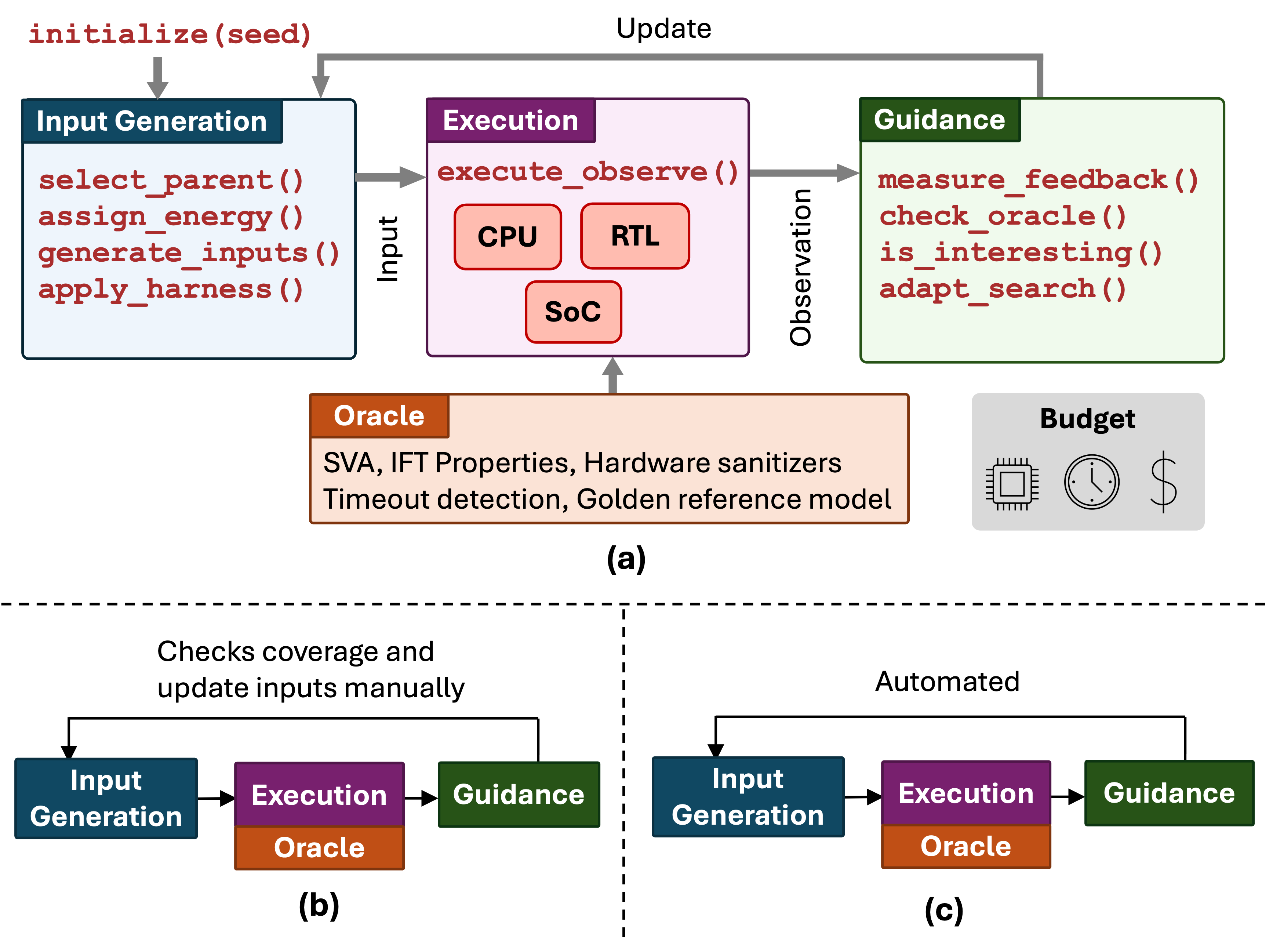}
    \caption{(a) Proposed framework: hardware fuzzing as budget-bounded search over input generation, execution, oracle, and guidance, with subcomponents from Algorithm~\ref{alg:generic-fuzzing}. (b) CRV relies on engineers to manually inspect coverage and update input generation; (c) fuzzing automates this verification loop through feedback-guided input generation. }
    \label{fig:framework}
\end{figure}

We define fuzzing as guided verification, where observations from an execution influence how inputs are generated for future executions.
As shown in Figure~\ref{fig:framework}, our framework begins with a stated \textbf{verification objective}, which is the behavior, requirement, verification-plan item, or security property that a campaign seeks to exercise or falsify. The same DUT can be tested in campaigns with different verification objectives. 
The verification objective drives every other choice in the campaign.

\subsection{Core Components of a Fuzzer}

We analyze a verification campaign through three components. 
The oracle determines which failures can be recognized, input generation determines which behavior is tested, and guidance determines how observations affect the subsequent input generation.
Together, these three determine what bugs or failures the fuzzer can find within a finite budget.

\paragraph{Oracle} determines whether an observed execution violates a checked specification.
It connects observations from the DUT to the design specification codified by the verification objective.
Some oracles check only the DUT’s externally visible outputs. 
Others also inspect internal state or sequences of events across multiple clock cycles.
In hardware verification, every class of failures we intend to find in a campaign must be explicitly defined in the oracle. 
Anything not checked remains invisible.

\paragraph{Guidance} determines which executions receive attention under a finite campaign budget.
The observation from an execution is turned into a \textbf{feedback} signal, which is used to guide the input generation.
The feedback can be either \textbf{coverage} over a set of observations, or a scalar function (\textbf{score}) computed from the observations.
The fuzzer generates inputs with the goal of either maximizing coverage or optimizing the score function.
For performance reasons, the feedback used for guidance is a lossy summary of a larger execution.
Useful guidance preserves distinctions that matter to the verification objective while remaining inexpensive enough to collect throughout the campaign.
If it rewards behavior unrelated to the objective, the fuzzer may make measurable ``progress'' without ever generating an input that will expose a failure.

\paragraph{Input generation} determines which tests can be expressed and which behaviors are practically observable.
It includes the input representation, seed corpus (initial inputs), generation and mutation mechanisms, and the policies that allocate generation effort.
Expressibility and reachability limit a campaign in different ways.
A representation may permit a test while the available seeds and transformations make it extremely unlikely to appear within the budget.
Conversely, a generator may produce a diverse set of tests efficiently while accidentally excluding a specific test required to expose a failure.
The \textbf{harness} maps a candidate from a fuzzer into an executable input to the DUT while preserving the timing, protocol, and system interactions needed by the campaign.
If the input language or harness cannot express a triggering action, guidance cannot help to reach it.

\subsection{Target Abstraction and Budget}
\label{sec:framework-abstraction}

\textbf{The target abstraction} constrains every component of the framework.
Whether the DUT is treated as an IP, CPU, or SoC changes what observations can be used for guidance, which input representation is useful, which checks are practical, and how much each execution costs.
It also changes how the three components interact.
Limited visibility constrains both the oracle and the feedback, while limited controllability constrains the harness and input representation.
Execution cost limits how many tests the campaign can run.

A finite \textbf{budget} forces tradeoffs among these choices.
A test that is technically reachable may still be irrelevant in practice if generating or executing it uses up the budget before the violation can be observed.
In contrast, a more expensive feedback mechanism may increase per-execution time while reducing the number of executions needed to find a failure, potentially reducing total campaign time.
Therefore, time, compute, memory, and human effort can all form part of this budget.

Together, the oracle, guidance, and input generation determine which failures a campaign can find within its budget. 
This is what we mean by \emph{you find what you seek}.

%% file: background.tex
\section{Background}
\label{sec:background}

The framework in Section~\ref{sec:framework} describes fuzzing in terms of its objective, oracle, guidance, input generation, target abstraction, and budget.
This section applies that vocabulary to software fuzzing, a generalized hardware-fuzzing loop, and established dynamic hardware verification practices, including CRV and property-driven security verification.

Hardware verification utilizes both static and dynamic techniques.
Formal verification, a static technique, is considered the gold standard in verification, but it does not scale for complex designs such as large SoCs and CPUs.
Dynamic verification executes concrete tests against a design in simulation, emulation, or physical hardware and observes the resulting traces.
CRV is the most common form of dynamic hardware verification, in which inputs are generated randomly within constraints.
In contrast, hardware fuzzing is a newer approach that uses observations from earlier executions to guide the generation of future test inputs.
While this SoK focuses on dynamic campaigns, we include static analysis techniques when they are used to guide hardware fuzzing.

\subsection{From Software to Hardware Fuzzing}

Fuzzing is a self-guided verification technique in which inputs are generated based on feedback from prior executions.
Early software fuzzers did not use such guidance~\cite{sw_miller_1990}, but it is a defining feature of contemporary fuzzers~\cite{aflplusplus,sw_art_science}.
A fuzzer can use this feedback to seek new coverage, reduce a distance or cost function, or reach an intermediate state associated with a target property.
AFL++ is a representative software fuzzer, and its mutational loop roughly resembles Algorithm~\ref{alg:generic-fuzzing}~\cite{aflplusplus}.
Consider AFL++ fuzzing a PDF reader.
It starts from a small corpus of PDF files, selects one as a parent, mutates its bytes, and uses a harness to invoke the reader.
Lightweight instrumentation records the control-flow edges and bucketed edge counts activated by the execution.
A candidate enters the corpus if it activates a new edge or a new count bucket for a known edge.
AFL++ also records execution time, input size, path frequency, and favored status to prioritize parents and assign mutation effort.

However, software fuzzing techniques do not directly transfer to hardware.
While software fuzzers can still encounter silent failures and structured inputs, these failures take different forms in hardware, as incorrect DUT behavior generally does not crash the simulator, and bugs may span numerous signals and clock cycles.
These differences affect: (i) the harness, which must translate a candidate into structured input over multiple clock cycles; (ii) the guidance, which must summarize relevant hardware behavior; and (iii) the oracle, which must apply explicit design- or security-specific checks.

\subsection{Anatomy of a Hardware Fuzzer}

\SetAlCapFnt{\small}
\SetAlCapNameFnt{\small}
\definecolor{algred}{RGB}{170,45,45}
\definecolor{algblue}{RGB}{35,85,150}
\newcommand{\op}[1]{%
  \textcolor{algred}{\bfseries\ttfamily\upshape #1}}
\SetKwFor{While}{\textcolor{algblue}{while}}{}{}
\SetKwFor{ForEach}{\textcolor{algblue}{for}}{}{}
\SetKwIF{If}{ElseIf}{Else}
  {\textcolor{algblue}{if}}{}
  {\textcolor{algblue}{else if}}
  {\textcolor{algblue}{else}}{}
\begin{algorithm}[t]
\caption{Feedback-guided Hardware Fuzzing as pseudocode. Individual fuzzers may omit or combine operations and do not necessarily implement every function shown.}
\label{alg:generic-fuzzing}
\footnotesize
\DontPrintSemicolon
\SetAlgoNoEnd
\ttfamily
seed: initial set of candidate inputs\;
corpus,feedback\_history,metadata = \op{initialize}(seed)\;
fail\_history = \{\}\;
\While{campaign budget remains}{
    parent = \op{select\_parent}(corpus, metadata)\;
    energy = \op{assign\_energy}(parent, metadata)\;
    candidates = \op{generate\_inputs}(parent, energy)\;
    \ForEach{candidate $\in$ candidates}{
        test\_vector = \op{apply\_harness}(candidate)\;
        observation=\op{execute\_observe}(test\_vector)\;
        feedback = \op{measure\_feedback}(observation)\;
        failure = \op{check\_oracle}(observation)\;
        \If{failure}{
            fail\_history.update(candidate,failure,feedback)
        }
        \If{\op{is\_interesting}(feedback)}{
            feedback\_history.update(feedback)\;
            corpus.append(candidate, observation)\;
        }
        metadata.update(parent,candidate,observation)\;
    }
    \op{adapt\_search}(corpus,feedback\_history,metadata)\;
}
\end{algorithm}

Algorithm~\ref{alg:generic-fuzzing} provides a high-level description of a hardware fuzzer and a common vocabulary for this SoK, while Figure~\ref{fig:framework} shows how its components interact.
The loop selects a retained input, generates and executes candidates, measures progress, saves failures, retains interesting inputs, and repeats.
Retained inputs form the \textbf{corpus}.
The selected input is the \textbf{parent}, and a newly generated input is a \textbf{candidate}.
A \textbf{harness} maps the candidate to executable DUT stimulus.
\textbf{Feedback} summarizes progress and guides future exploration, while the \textbf{oracle} defined in Section~\ref{sec:framework} independently checks whether the execution exposes a failure.

\textbf{initialize()} builds the corpus from external, random, or solver-derived seeds.
\textbf{select\_parent()} may traverse the corpus, sample uniformly or by weight, or rank inputs by target distance.
\textbf{assign\_energy()} fixes or adjusts mutation effort using metadata such as coverage, execution cost, distance, rarity, or past reward.
\textbf{generate\_inputs()} performs byte, bit, instruction, transaction, grammar-aware, structure-aware, or adversarial-action mutation.
It may also combine parents or add solver-derived or language-model-generated candidates.

\textbf{apply\_harness()} preserves the structure required by the target abstraction.
RTL/IP fuzzers map bits or bytes to pins over several cycles.
SoC/NoC fuzzers produce protocol transactions, firmware actions, or adversarial tampering, while CPU fuzzers produce instructions, initial memory, and interrupts.
\textbf{execute\_observe()} runs simulation or emulation on  a field programmable gate array (FPGA).
Fuzzers may use checkpoints, perform soft resets, or compare paired executions.

\textbf{measure\_feedback()} derives coverage or a scalar score from the observation.
Coverage may track control-flow-graph (CFG) edges, mux toggles, control registers, taint states, information-flow paths, or NoC transactions, while scores may estimate distance to a vulnerability condition.
\textbf{check\_oracle()} detects crashes, hangs, assertion or hyperproperty violations, differential mismatches, timing differences, taint leaks, and sanitizer findings.
\textbf{is\_interesting()} compares feedback with campaign history and retains candidates that expand coverage or improve the best-known target distance, vulnerability cost, or another score.
Some fuzzers perform no adaptation, while others use \textbf{adapt\_search()} to respond to stagnation by changing mutations, combining inputs, resetting state, or invoking symbolic or formal generation.

\subsection{CRV vs. Hardware Fuzzing}
\label{sec:verification-objectives}

{

\SetInd{0.25em}{0.5em} 
\definecolor{algred}{RGB}{170,45,45}
\definecolor{algblue}{RGB}{35,85,150}
\definecolor{algblack}{RGB}{0,0,0}
\definecolor{algcomment}{RGB}{85,140,95}
\newcommand{\component}[1]{\textcolor{algblack}{\bfseries\ttfamily\upshape #1}}
\newcommand{\algcomment}[1]{\textcolor{algcomment}{\bfseries #1}}
\SetAlCapFnt{\small}
\SetAlCapNameFnt{\small}
\begin{algorithm}[t]
\caption{
Constrained-Random Verification of an OpenTitan Access-Control IP as pseudocode. Engineer-defined constraints generate structured TL-UL requests, functional coverage measures completeness, and a golden reference model, scoreboard, and assertions form the oracle.
}
\label{alg:crv}
\footnotesize
\DontPrintSemicolon
\SetAlgoNoEnd
\ttfamily

ac = \component{ConfigGen\_AccessControl}()\;

ac.ranges.\op{constraint}(\algcomment{\{valid address ranges\}})\;
ac.permissions.\op{constraint}(\ldots)\;
\BlankLine

gen = \component{InputGen\_TLUL}()\;
gen.addr.\op{constraint}(ac.ranges,\algcomment{\{higher weights around selected range\}})\;
\BlankLine

cov=\component{Coverage}(ac.ranges,ac.permissions,ac.decisions)\;
cov.bins = \op{cross}(ranges, permissions, decisions)\;
cov.bins.\op{illegal\_bins}(\algcomment{\{excluded bins\}})\;
\BlankLine

orc = \component{Oracle}(GRM(),Monitor(),Scoreboard(),Asserts())\;
\While{total campaign budget remains}{
    seed = \op{next\_seed}()\;
    dut.config = ac.\op{randomize}(seed, permissions)\;
    sequence = gen.\op{randomize}(seed,addr,data,command)\;

    \ForEach{tlul\_req $\in$ sequence}{
        observation = \op{execute\_observe}(tlul\_req)\;
        expected = orc.grm(tlul\_req, dut.config)\;

        cov.\op{sample}(tlul\_req, expected, observation)\;
        orc.\op{run\_scoreboard}(expected, observation)\;
        orc.\op{assert}(\algcomment{\{completes without timeout\}})\;
        orc.\op{assert}(\algcomment{\{denied read $\Rightarrow$ data = 0\}})\;
    }
    \BlankLine
    \textit{\op{//manually performed by engineers:}} \algcomment{\{\;
    \Indp
    inspect coverage for gaps\;
    update constraints, tests, gaps\;
    \Indm
    \}\;}
}
\end{algorithm}
}

Hardware security verification combines functional verification of security-critical components with requirements derived from a threat model.
Functional plans encode expected behavior through tests, covergroups, scoreboards, assertions, and reference models.
Threat models identify assets, adversary capabilities, trust boundaries, and permitted behavior, then derive requirements for isolation, confidentiality, integrity, and availability.
Both are needed because functional bugs can become vulnerabilities, while functional correctness alone does not establish resistance to an adversary.

CRV is the established approach for exercising functional verification plans and is typically implemented with standardized universal verification methodology (UVM) in SystemVerilog.
Algorithm~\ref{alg:crv} distills the verification environment for OpenTitan's \texttt{ac\_range\_check} IP~\cite{opentitan_ac_range_check_dv}.
OpenTitan is a root of trust widely adopted in consumer electronics whose per-IP environments include test plans, constrained sequences, drivers, monitors, scoreboards, assertions, and coverage models~\cite{opentitan_dv_methodology}.
The IP enforces a Register Access Control List (RACL) role policy by checking TL-UL requests against configurable address ranges and read, write, and execute permissions~\cite{opentitan_ac_range_check}.
Algorithm~\ref{alg:crv} contrasts its constrained input generation, functional coverage, and oracle with the hardware-fuzzing loop in Algorithm~\ref{alg:generic-fuzzing}.
CRV samples engineer-defined constraints and relies on engineers to revise campaigns, while fuzzing uses feedback from earlier executions to guide later inputs.

Random inputs in CRV remain structured and need not follow uniform distributions.
In this example, constraints ensure valid access-control ranges and permissions.
For each seed, \texttt{randomize} produces a configuration and a TL-UL sequence that preserves request structure across signals and cycles while varying \texttt{addr}, \texttt{data}, and \texttt{command}.
The address constraint assigns 98\% of its weight to a selected range and nearby addresses, with 2\% assigned to the extremes of the address space.
Enabled read and write permissions receive twice the weight of disabled permissions.

Functional coverage records a \texttt{cross} (combination) of ranges, permissions, and decisions, capturing interactions that coverage of each attribute alone would miss.
It excludes unnecessary combinations to make coverage meaningful, and \texttt{cov.sample} records the observed interactions.
Engineers inspect remaining gaps after a campaign and add tests, revise constraints, or document unreachable cases.
The monitor, golden reference model, scoreboard, and assertions form the oracle.
The monitor observes each transaction, the model computes the expected response, and the scoreboard compares and records expected and observed behavior.
Assertions check same-cycle or temporal relationships.
In this example, they detect hangs and data leakage from unauthorized reads.
Functional coverage measures campaign completeness, while the scoreboard and assertions detect failures.

At its simplest, CRV resembles early software fuzzing, which generated random inputs and treated crashes or hangs as failures without using execution feedback to guide later inputs~\cite{sw_miller_1990,sw_art_science}.
Industrial CRV is far more mature than early software fuzzing, but its feedback loop remains manual. 
Modern fuzzers automate this loop through guided input generation, which we use to distinguish hardware fuzzing from CRV.
This boundary is shifting. 
A technique in a 2024 Synopsys patent guides constraint solving using coverage history, while Cadence Xcelium ML and Siemens Questa One Sim CX infer relationships between stimulus and coverage to accelerate coverage closure~\cite{synopsys_patent_2024,cadence_xcelium_ml,siemens_questa_sim_cx}.
CRV and fuzzing lie on a continuum.
Their convergence motivates hardware fuzzing to combine lessons from software fuzzing with mature CRV practices under hardware-specific objectives and constraints.
This operational view provides context for the Oracle, Guidance, and Input Generation sections that follow.

%% file: oracle.tex
\section{Oracle}
\label{sec:oracle}

Hardware fuzzers differ not only in how they generate inputs and guide exploration, but also in how they determine that an execution is erroneous.
At its core, an \textit{oracle} is the mechanism that determines whether the DUT deviates from an expected behavior.
The expected behavior may be represented explicitly as a property assertion or implicitly through a reference model or a comparison between executions.
To formalize failure detection across these different fuzzing setups, we use a unified vocabulary based on three concepts: \textit{property}, \textit{assertion}, and \textit{failure}.
A \textit{property} is a computable expression that encodes a particular expected behavior.
An \textit{assertion} is a mechanism that evaluates whether a property holds and reports a failure when it does not.
We use \textit{failure} to denote the detection of a deviation between the implemented and expected behavior, regardless of whether that deviation is identified through an assertion, a reference-model mismatch, a differential comparison, or another checking mechanism.
This practical definition captures the failure-detection mechanisms used by hardware fuzzers in our survey without restricting the oracle to a particular formal representation of expected behavior.

\subsection{Oracle Mechanisms and Blind Spots}

The way expected behavior is represented determines what an oracle can observe and, consequently, which classes of erroneous behavior a fuzzer can detect.
For example, an architectural reference model can identify deviations from the architectural specification of a processor, but may not expose microarchitectural or information-flow violations that do not manifest in architectural state~\cite{geier2025,symbfuzz,milesan}.
Other fuzzers therefore rely on specialized models, comparisons between multiple executions, or explicitly specified properties and verification monitors.
Across the surveyed fuzzers, these mechanisms can be organized into four broad oracle families, summarized in Table~\ref{tab:oracle}.
The first uses an architectural reference model to compare the DUT against an ISA-level representation of expected behavior.
The second uses specialized models to capture microarchitectural, security, timing, or information-flow behavior that is not represented by architectural state.
The third identifies failures through differential or relational comparisons between executions or states within the DUT.
The fourth relies on explicitly specified properties and verification infrastructure, including assertions, monitors, scoreboards, and sanitizers.
These families describe how expected behavior is represented and are not mutually exclusive; an oracle may combine several complementary mechanisms.

For each executed test, the oracle determines whether the observed behavior violates the specification.
Assertions and protocol monitors enforce local, temporal, and interface rules; timeout checks detect hangs; and scoreboards and reference models detect departures from expected execution~\cite{fuzz_hw_like_sw}.
Sanitizers look for recurring weakness patterns~\cite{fuzzitizer}.
Hyperproperties can encode confidentiality, integrity, and non-interference requirements; hardware fuzzers can check them through self-composition, miter, or information-flow tracking mechanisms~\cite{hyperfuzzing,opentitan_ift}.
Any failure undetected by the oracle remains invisible to the verification campaign~\cite{isadora,fuzzitizer}.

The oracle must likewise match the target and threat model.
An IP may use local assertions, protocol checks, or a block-level reference model; a CPU may use architectural differential testing; and an SoC may require properties spanning firmware, interconnects, peripherals, and security state~\cite{difuzzrtl,hyperfuzzing,fuzzitizer}.
A local assertion does not establish an SoC-wide security invariant~\cite{fuzzitizer}.

\begin{table}
\centering
\setlength{\tabcolsep}{3pt}
\caption{
Oracle mechanisms across the surveyed fuzzers.
}
\label{tab:oracle}
\footnotesize
\include{tables/oracle/classification}
\vspace{-2.5em}
\end{table}

\subsection{Failure Validation and Triage}

An oracle report identifies a potential failure, not necessarily a confirmed bug or its root cause~\cite{genhuzz,expect}.
When a failure occurs, the input sequence that generated it must be analyzed alongside the DUT's RTL, specification, and properties, to determine what went wrong and where a fix must be applied~\cite{opentitan_ift}.
This analysis becomes more difficult when a campaign continues after the first failure and produces many reports, including duplicate manifestations of the same bug~\cite{genhuzz}.
Therefore, the reports must be reproduced, minimized, grouped, and validated before they become useful findings~\cite{genhuzz, cascade}.

Assigning a relative priority to validated failures requires assessing their impact on functionality, security, or safety and the scope of the repair~\cite{opentitan_ift}.
Their frequency, reproducibility, and analysis effort provide additional operational criteria.
Fuzzers may assist this process using information they already collect.
Automated input reduction can preserve the bug-triggering state while making long programs tractable to analyze, and signatures derived from execution traces can group duplicate reports~\cite{cascade,genhuzz}.
Future hardware-fuzzing papers should report the manual effort spent, the triage or severity-scoring method employed, and the post-processing needed to make the fuzzer's output useful.

Google's OSS-Fuzz project shows that the fuzzing loop is only one part of deploying fuzzing at scale.
Its largely automatic failure detection reports crashes, sanitizer findings, timeouts, and out-of-memory executions, while ClusterFuzz provides continuous fuzzing infrastructure~\cite{oss_fuzz}.
These reusable partial oracles require little target-specific specification.
Target project maintainers normally maintain recommended seed corpora and dictionaries where applicable~\cite{oss_fuzz}.
\begin{takeaway}
An oracle only detects failures that are both observable and explicitly checked.
Its findings should be validated and prioritized before being treated as bugs.
\end{takeaway}

%% file: tables/oracle/classification.tex

\begin{tabularx}{\linewidth}{
    @{}
    >{\raggedright\arraybackslash}p{0.20\linewidth}
    >{\raggedright\arraybackslash}p{0.50\linewidth}
    >{\raggedright\arraybackslash}X
    @{}
}
    \toprule
    \textbf{Oracle mechanism} & \textbf{Expected behavior / comparison} & \textbf{Fuzzers} \\
    \midrule
    Reference-model comparison (21)
    & DUT architectural, functional, or transaction-level behavior checked against a separate reference model or model-derived expected result.
    & \cite{difuzzrtl,processorfuzz,morfuzz,hypfuzz,thehuzz,mabfuzz,chatfuzz,pathfuzz,cascade,genfuzz,divefuzz,goldenfuzz,refuzz,rlfuzz,simfuzz,turbofuzz,psofuzz,genhuzz,fuzz_hw_like_sw,nocfuzzer,presifuzz} \\
    \addlinespace
    Specialized semantic / cost model (8)
    & A domain-specific execution, leakage, information-flow, contract, or vulnerability-cost model supplies the detection semantics.
    & \cite{geier2025,introspectre,milesan,phantomtrails,taintfuzzer,socfuzzer,dejavuzz,formalfuzzer} \\
    \addlinespace
    Differential / relational (11)
    & Paired executions, implementations, states, timing observations, or information-flow views are compared; ordinary DUT-versus-golden functional checking is listed above.
    & \cite{specdoctor,specure,sigfuzz,riscover,synfuzz,hyperfuzzing,sonar,whisperfuzz,geier2025,milesan,dejavuzz} \\
    \addlinespace
    Property / runtime checker (13)
    & An assertion, contract, security/protocol predicate, scoreboard, generated sanitizer rule, or expected-termination check evaluates each fuzzed execution.
    & \cite{symbfuzz,fuzz_hw_like_sw,fuzzitizer,interconfuzz,nocfuzzer,fuzzwiz,spinalfuzz,vgf,surgefuzz,genfuzz,hyperfuzzing,geier2025,cascade} \\
    \bottomrule
\end{tabularx}

\vspace{0.35em}
\parbox{\linewidth}{
\emph{Scope note.}
The four rows classify implemented failure-detection mechanisms and contain 53 non-exclusive placements covering 44 of the 52 surveyed frameworks.
BugsBunny and RFUZZ delegate relevance or failure detection to a user-defined target or external checker~\cite{bugsbunny,rfuzz}.
FMTC, TargetFuzz, DirectFuzz, FuSS, ProFuzz, and RTLFuzzLab report reachability or coverage but do not implement a pass/fail oracle~\cite{fmtc,targetfuzz,directfuzz,fuss,profuzz,rtlfuzzlab}.
}

%% file: guidance.tex
\section{Guidance}
\label{sec:guidance}

\begin{table*}[t]
\centering
\footnotesize
\setlength{\tabcolsep}{3pt}
\renewcommand{\arraystretch}{1.08}
\caption{
Guidance across 52 hardware fuzzers.
The \emph{type} of feedback is either \textbf{C}overage over a set of source observations to be expanded, or a numeric \textbf{S}core function to be optimized.
Counts are non-exclusive and fuzzers appear in every applicable row.
}
\label{tab:guidance-summary}
\begin{tabular}{@{}p{0.10\textwidth}p{0.04\textwidth}p{0.83\textwidth}@{}}
\hline
\textbf{Source} & \textbf{Type} & \textbf{Details of the guidance mechanism} \\
\hline

\multirow{2}{0.10\textwidth}{Implementation structure / state (32)}
& C (24)
& RTL or Verilator control flow~\cite{fuzz_hw_like_sw,spinalfuzz,fuzzwiz,symbfuzz,interconfuzz,phantomtrails};
mux toggle~\cite{rfuzz,rtlfuzzlab};
full-mux toggle~\cite{rtlfuzzlab,fmtc};
statement and expression~\cite{thehuzz};
line~\cite{rtlfuzzlab,nocfuzzer};
branch~\cite{fuss,thehuzz,hypfuzz,whisperfuzz,nocfuzzer};
condition and FSM~\cite{thehuzz,hypfuzz,whisperfuzz};
toggle~\cite{fuss,thehuzz,pathfuzz};
VCS code-coverage points~\cite{presifuzz,fuzzitizer};
per-cycle control-register state~\cite{difuzzrtl,morfuzz,turbofuzz,sigfuzz,bugsbunny};
architectural CSR transitions~\cite{processorfuzz};
netlist cell toggle and expression~\cite{synfuzz} \\
& S (8)
& structural-coverage-derived genetic-algorithm fitness~\cite{genfuzz};
particle-swarm optimization~\cite{psofuzz};
multi-armed and contextual-bandit reward~\cite{mabfuzz,refuzz};
language-model (LM) reinforcement learning (RL)~\cite{chatfuzz,genhuzz};
deep-RL reward~\cite{rlfuzz};
and LM tuning~\cite{goldenfuzz} \\
\hline

\multirow{2}{0.10\textwidth}{Selected region of design (6)}
& C (4)
& target nets~\cite{profuzz};
target-module mux activity~\cite{directfuzz};
changes on signal close to property~\cite{vgf};
ancestor-register state coverage~\cite{surgefuzz} \\
& S (2)
& distance from covered muxes to the target module sets energy~\cite{directfuzz};
target-signal dependency distance sets seed priority~\cite{bugsbunny} \\
\hline

\multirow{2}{0.10\textwidth}{User-defined behavior (3)}
& C (1)
& UVM functional covergroups for NoC~\cite{nocfuzzer} \\
& S (2)
& temporal density of an annotated event assigns mutation effort~\cite{surgefuzz};
packet waiting time~\cite{nocfuzzer} \\
\hline

\multirow{2}{0.10\textwidth}{Threat model \\ or security property (12)}
& C (6)
& adversary-action bigrams and fault-read events~\cite{hyperfuzzing};
secret-propagation taint matrix~\cite{dejavuzz};
self-composition divergence~\cite{geier2025};
toggles on statically found leakage paths within speculative window~\cite{specure};
taint propagation~\cite{phantomtrails};
opcode/rollback-reason pair~\cite{specdoctor} \\
& S (6)
& vulnerability cost function~\cite{socfuzzer,formalfuzzer};
taint and target-output proximity cost~\cite{taintfuzzer};
transient-window size, after a new trigger cause--opcode pair~\cite{specdoctor};
request interval at a contention point~\cite{sonar};
rarity-weighted self-composition-deviation prioritization~\cite{geier2025} \\
\hline

Program (1)
& S (1)
& deviation of executed opcode frequencies from the expected distribution rebalances opcode selection in later rounds~\cite{divefuzz} \\
\hline

Other (6)
& -
& No execution feedback; generation uses static/formal analysis, execution models, historical data, or rules.~\cite{targetfuzz,cascade,milesan,introspectre,riscover,simfuzz}\\
\hline
\end{tabular}
\end{table*}

Guidance determines how a fuzzer spends a finite campaign budget.
Feedback from earlier executions may cause the fuzzer to retain a candidate, prioritize a parent, assign more mutation effort, or change its search strategy.
Since guidance influences input generation, the boundary between the two can be blurred.
We therefore use \emph{guidance} for what behavior receives attention and \emph{input generation} for which candidates can be produced and how they are applied to the DUT.

A dynamic campaign executes only a small fraction of possible tests.
Therefore, coverage is used to assess completeness and guide exploration, but it remains a lossy proxy for progress rather than a verification result.
CRV uses structural coverage as a sanity check and relies primarily on functional coverage from the verification plan, whose covergroups can encode temporal, protocol-specific, and cross-event behavior.
NoCFuzzer illustrates the tradeoff: CRV closed an easy router covergroup faster, whereas fuzzing closed harder code and functional targets that CRV did not reach within the budget~\cite{nocfuzzer}.
Guidance is beneficial only when it reduces search effort enough to offset the cost of feedback.

Table~\ref{tab:guidance-summary} summarizes execution-derived guidance across the surveyed fuzzers.
The \emph{source} records what is observed, while the \emph{type} records whether those observations become coverage or a numeric score.
The same feedback type can carry different meanings depending on whether it comes from implementation structure, a selected region, a verification plan, or a threat model.
The fuzzers in the \emph{Other} row do not adapt their search using feedback from earlier DUT executions.
Instead, their user-selected targets, static or formal analysis, execution models, historical data, or construction rules shape input generation~\cite{targetfuzz,cascade,milesan,introspectre,riscover,simfuzz}.

AFL helped make feedback-guided mutational fuzzing practical and widely adopted by popularizing approximate CFG-edge coverage and bucketed hit counts as inexpensive search proxies~\cite{sw_art_science,aflplusplus}.
Hardware can reuse this design pattern, but the useful feedback changes with the abstraction level.
Fine-grained signals such as mux toggles may guide a fuzzer on an isolated IP but become noisy or expensive at CPU and SoC scale.
At those levels, guidance may track control and status register (CSR) transitions, instruction sequences, NoC transactions, or functional covergroups~\cite{hyperfuzzing,processorfuzz,nocfuzzer}.

\subsection{Source and Types of Feedback}

The most common source of feedback is the implementation of the design itself.
Line, branch, condition, toggle, mux, and register-state coverage can be derived automatically from the RTL or netlist of the design.
They require less target-specific user effort than a functional coverage model or a security property, but show only that a test exercised new logic.
For example, the select pins of four 2:1 muxes can go from \texttt{0000} $\rightarrow$ \texttt{1111} $\rightarrow$ \texttt{0000} and achieve 100\% mux-toggle coverage in a single, short trace.
Yet four independent select pins create up to $2^{4 \times 10} \approx 1$ trillion traces over ten cycles, and mux-toggle coverage preserves neither event order nor protocol context.

Coverage feedback is constructed by choosing observable values and defining coverage items over them.
Instrumentation monitors those values, and the fuzzer retains a candidate when it contributes an item absent from the accumulated history.
RFUZZ's mux-select activity, NoCFuzzer's UVM covergroups, and HyperFuzzing's instruction, NoC-access, and fault-read events all use novelty, but preserve different information about an execution~\cite{rfuzz,nocfuzzer,hyperfuzzing}.

Not all hardware fuzzers are coverage-guided.
Directed fuzzers instead compute a numeric score such as target distance, security cost, estimated influence on an asset, or contention interval~\cite{bugsbunny,socfuzzer,taintfuzzer,formalfuzzer,sonar}.
NoCFuzzer prefers inputs that increase packet waiting time and move toward starvation, while TaintFuzzer's cost includes inferred input influence on outputs and proximity to asset leakage~\cite{nocfuzzer,taintfuzzer}.
Such scores can measure directed progress, but they still remain lossy.

Coverage and score-based feedback can coexist.
BugsBunny first uses register coverage to discard uninteresting tests and computes target distance for survivors~\cite{bugsbunny}.
DirectFuzz similarly uses target-module coverage to prioritize corpus entries and a distance score to assign mutation energy~\cite{directfuzz}.
This can avoid computing an expensive score for every test.
The score's optimizer is a separate choice. 
GenFuzz and PSOFuzz apply genetic and particle-swarm search to structural coverage without changing the feedback signal~\cite{genfuzz,psofuzz}.

Feedback derived from a security specification is also guidance.
HyperFuzzing's event bigrams are intended to lead the search toward adversarially affected behavior, while its HyperPLTL oracle determines whether the paired executions violate a property~\cite{hyperfuzzing}.
An intermediate taint state can likewise guide a fuzzer to expose an asset-to-sink leakage path.

\subsection{Target Selection}
\label{sec:target-selection}

\begin{figure}
    \centering
    \includegraphics[width=1\linewidth]{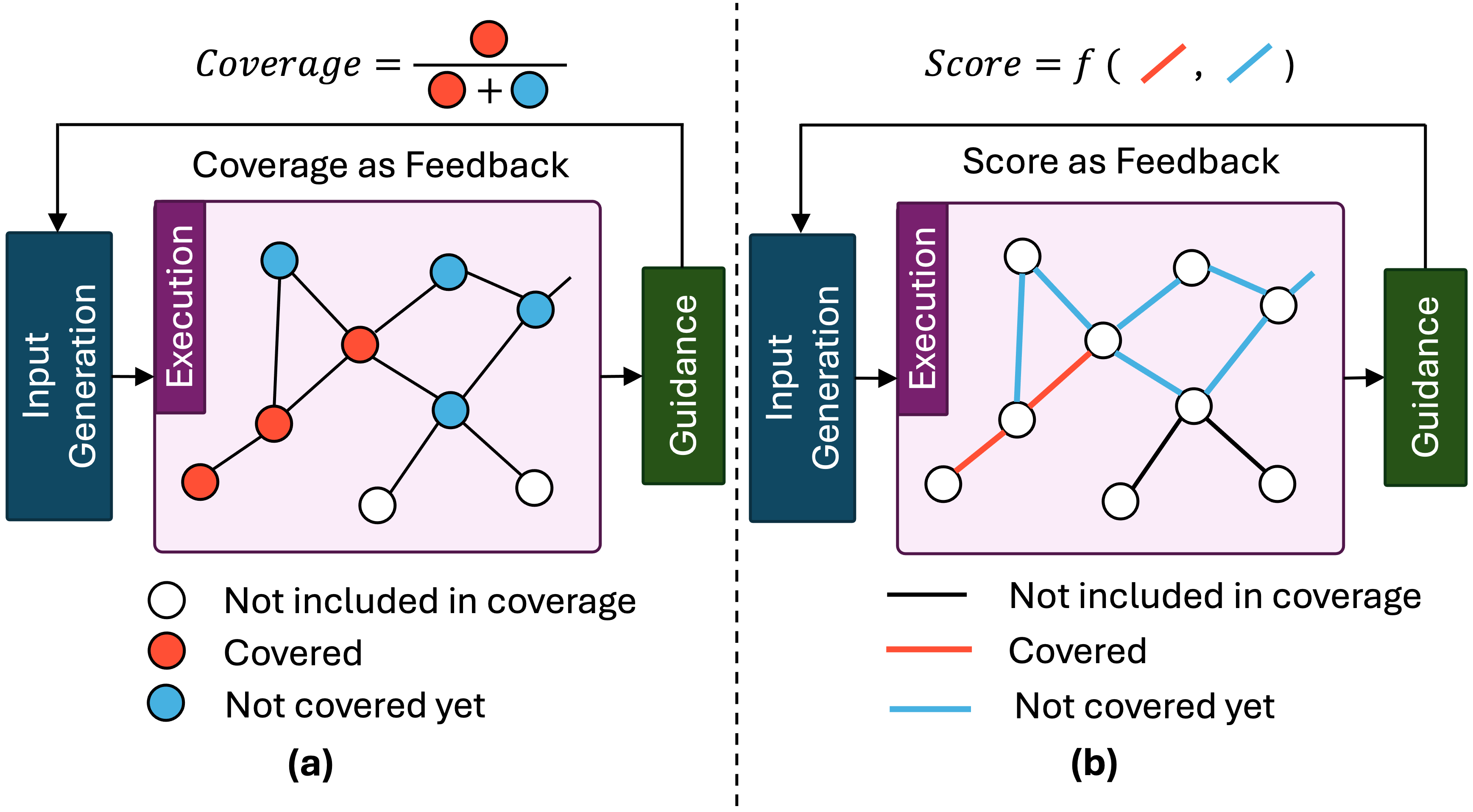}
    \caption{Types of guidance (a) Coverage rewards new coverpoints or register states. (b) A score function $f()$ measures progress toward a target, e.g., asset-to-output leakage.}
    \label{fig:cov_score}
\end{figure}
Adopting directed guidance for fuzzing requires defining the specific target of the campaign.
An engineer must state the verification goal, a user or algorithm must select a target signal or region, and the input generator must produce an input that activates it.
We treat the first two as guidance and the last as input generation.
An automatic solver can construct an input for a target, but it cannot determine whether that target represents the user's verification intent.

For security verification, the target normally originates in a threat model or property.
VGF automates part of this translation by deriving monitored signals from a user-supplied security property~\cite{vgf}.
ProFuzz  selects nets manually, randomly, or by scoring fan-in, fan-out, and Shannon entropy in a netlist hypergraph~\cite{profuzz}. Automatic targets remain structural unless connected to a verification plan or threat model.

\subsection{Alignment Across the Guidance Chain}

A guidance signal is the end of a chain of choices.
The verification objective is translated into a target behavior, the behavior into observable values, and those observations into coverage or a score.
A search policy or an optimizer then uses the resulting signal to retain candidates, select parents, or assign effort.
Each mapping can discard information.
Functional coverpoints begin from verification-plan behavior, and security-derived scores begin from a property, asset, or vulnerability condition~\cite{nocfuzzer,vgf,taintfuzzer}.
Structural feedback begins from readily observable implementation activity and assumes that new activity correlates with the objective~\cite{rfuzz}.
Thus, coverage over a functional coverpoint may align better than a distance score to a structurally selected module, while a score derived from a security condition may align better than structural coverage.
No optimizer can recover distinctions discarded earlier in the chain.
\begin{takeaway}
Feedback is a lossy proxy and is not evidence that a design is bug-free.
Guidance should be designed backward from the verification objective through target behavior, observation, feedback, and search policy.
\end{takeaway}

%% file: input_gen.tex
\section{Input Generation}
\label{sec:input-generation}

\begin{table*}[t]
\centering
\footnotesize
\setlength{\tabcolsep}{3pt}
\renewcommand{\arraystretch}{1.08}
\caption{
Input generation across 52 hardware fuzzers.
The first column is how a fuzzer treats the target.
For example, a fuzzer may treat a CPU as an IP.
The second column defines how an input is generated: \textbf{(M)}utation of a parent, \textbf{(C)}onstruction without a parent, and \textbf{(S)}olving for target.
Counts are non-exclusive and fuzzers with composite mechanisms appear in multiple rows.
}
\label{tab:input-gen-summary}
\begin{tabular}{@{}p{0.04\textwidth}p{0.035\textwidth}p{0.90\textwidth}@{}}
\hline
\textbf{Target} & \textbf{Gen.} & \textbf{Details of the input-generation mechanism} \\
\hline

\multirow{2}{0.04\textwidth}{IP (10)}
& M(9)
& bytes $\rightarrow$ pins $\times$ cycles~\cite{rfuzz,directfuzz,spinalfuzz,vgf,fuzzwiz,rtlfuzzlab,fmtc};
ATPG-seed bit-vector mutation~\cite{profuzz};
packet (per-cycle pin values)-level mutation~\cite{bugsbunny}; \\
& S(3)
& Inputs by SAT solving for target-node states with Hamming-distance constraint for diversity~\cite{targetfuzz};
merged ATPG activation patterns for target nets~\cite{profuzz};
recorded state of a partially toggled mux initializes Z3-backed completion~\cite{fmtc} \\
\hline

\multirow{3}{0.04\textwidth}{SoC / NoC (9)}
& M(7)
& 
SoC-harness is firmware in FPGA emulation~\cite{socfuzzer,taintfuzzer,formalfuzzer};
same, with manual constraints from formal traces~\cite{formalfuzzer};
bytes configure firmware/NoC/memory flip tamperers~\cite{hyperfuzzing};
bytes $\rightarrow$ UVM sequences~\cite{nocfuzzer};
minimized IP bus transactions $\rightarrow$ SoC program seeds $\rightarrow$ mutation~\cite{fuzzitizer};
mutation on opcode/operand fields of a precompiled benchmark program~\cite{fuss} \\
& C(2)
& CRV-style UVM randomization of input vectors~\cite{symbfuzz} or packet fields~\cite{interconfuzz}, with constraints tightened when coverage stalls \\
& S(3)
& a minimal suffix solved from the fuzzing trajectory at a plateau~\cite{fuss};
checkpoint-derived constraints that restrict later UVM randomization~\cite{symbfuzz,interconfuzz} \\
\hline

\multirow{3}{0.04\textwidth}{CPU (31)}
& M(23)
& ISA-instruction or program-IR edits~\cite{thehuzz,presifuzz,phantomtrails,genfuzz,mabfuzz,refuzz,rlfuzz,whisperfuzz,simfuzz};
instructions, interrupts, and addresses set by program execution~\cite{difuzzrtl,processorfuzz,sigfuzz,surgefuzz,geier2025,morfuzz,pathfuzz};
testcase-template/transient-seed mutation~\cite{sonar,specure};
editing tests they first construct or solve~\cite{psofuzz,specdoctor,turbofuzz,dejavuzz,hypfuzz};\\
& C(12)
& programs emitted by a language model~\cite{chatfuzz,genhuzz,goldenfuzz};
sampled from fixed or learned instruction distributions~\cite{riscover,psofuzz,specdoctor,turbofuzz};
constructed under an ISS or execution model that supplies operands, prerequisites, page tables, or entangled control and data flow~\cite{cascade,milesan,introspectre,divefuzz,dejavuzz} \\
& S(1)
& a model checker solves an uncovered point into an executable test~\cite{hypfuzz} \\
\hline

Misc(3)
& M(3)
& bytes $\rightarrow$ TileLink transactions driving an IP~\cite{fuzz_hw_like_sw, rtlfuzzlab};
netlist inputs labelled control \& data by user and mutated separately~\cite{synfuzz} \\
\hline
\end{tabular}
\end{table*}

Input generation determines which tests the fuzzer can produce given the feedback from guidance.
A bug whose triggering input is never generated will stay undetected.
Input generation includes the candidate representation, seed corpus, mutation or construction operations, solver assistance, allocation of generation effort, and the harness that applies a candidate to the DUT.
Table~\ref{tab:input-gen-summary} groups these mechanisms into mutation, construction without a parent, and solving backward from a target.

\subsection{Input Grammar and Harness}

Grammar is the structure that defines how the fuzzer's inputs are interpreted by the DUT.
The same byte string generated by mutation can be sliced into pin values, decoded into a transaction sequence, interpreted as a program, or consumed as commands for an adversarial tamperer.
The harness is the interface that applies the grammar to generate inputs and is therefore part of the input generator.
Industry-standard CRV has counterparts to both: the OpenTitan example in Algorithm~\ref{alg:crv} preserves the TL-UL transaction grammar and protocol, while its UVM driver serves a harness-like role by applying generated transactions to the DUT.

The generator must balance validity with flexibility.
Too little structure wastes executions on invalid or benign inputs, while too much can exclude the bug-prone or adversarial interactions the campaign is intended to test.
A transaction generator fuzzing a crossbar without concurrent requests cannot expose contention failures. A CPU-specific input generator without interrupts cannot expose a bug that requires one.

At the IP level, a harness can typically map candidate bytes to pin values over clock cycles or to selected transaction fields.
CPUs instead require programs together with execution state such as initial memory, privilege modes, and interrupts.
At SoC and NoC scale, the generator must coordinate processors, firmware, interconnects, peripherals, and security state over longer executions.
Such harnesses translate candidates into programs, firmware actions, packets, or UVM transactions while preserving ordering, handshakes, concurrency, and responses~\cite{fuzz_hw_like_sw,nocfuzzer,fuzzitizer}.
Firmware alone may not express malicious NoC transactions or memory faults. The grammar must include explicit threat-model entry points~\cite{hyperfuzzing}.

Software fuzzing encountered and overcame similar tension between validity and flexibility.
Production fuzzers now use target-specific corpora, dictionaries, parser-aware harnesses, and custom mutators. 
Csmith and syzkaller similarly encode domain knowledge for C programs and system-call sequences~\cite{oss_fuzz,sw_compiler_bugs,sw_art_science}.
Hardware can reuse an ISA across CPUs, but IP and SoC harnesses remain protocol- or design-specific.

RFUZZ maps bytes directly to primary-input values over successive cycles~\cite{rfuzz}.
This requires little design knowledge but does not preserve bus transactions; Fuzzing Hardware Like Software instead decodes bytes into TileLink transactions, and NoCFuzzer uses a user-written grammar to construct UVM sequences across NoC agents~\cite{fuzz_hw_like_sw,nocfuzzer}.
NoCFuzzer also shows why harness design is critical. 
A mapping in which each mutated byte affected multiple NoC queues performed worse than random testing~\cite{nocfuzzer}.

At the SoC level, the grammar also reflects how the adversary is modeled.
Transaction-aware harnesses avoid protocol-invalid tests, but may not express attacks that deliberately break protocol, such as those explored by eXpect and Xray~\cite{expect,xray}.
HyperFuzzing uses tamper functions for untrusted-firmware instructions, NoC writes, and memory bit flips~\cite{hyperfuzzing}.
SoCFuzzer and TaintFuzzer instead run target-specific C executables on the SoC and mutate their inputs~\cite{socfuzzer,taintfuzzer}.
The harness, grammar, seeds, and encoded domain knowledge should therefore be versioned campaign artifacts.

\subsection{Candidate Generation Mechanisms}

Surveyed fuzzers mutate parents, construct inputs without parents, solve backward from targets, or combine these mechanisms. 
Mutation engines may transfer across targets, but a complete generator may not because input grammars differ.

\subsubsection{Mutation and Construction}

Parent-derived mutation applies deterministic bit flips or arithmetic changes and nondeterministic ``havoc'' operations such as insertion, deletion, and overwriting. 
Splice or recombination operations draw from multiple parents~\cite{rfuzz,aflplusplus}.

Inputs can also be constructed without being derived from a parent.
SymbFuzz randomizes UVM input vectors and InterConFuzz randomizes packet fields while preserving sequence structure~\cite{symbfuzz,interconfuzz}.
After coverage stagnates, both add solver-derived constraints to redirect generation.
Constructed or solved inputs can then become parents for mutation.

\subsubsection{Construction using Language Models}

Several recent CPU fuzzers use generative pre-trained transformer (GPT)-like language models to construct RISC-V programs and fine-tune their generation using execution feedback without mutating a parent.
ChatFuzz pretrains on Linux-kernel-derived RISC-V code and uses disassembler validity and RTL coverage as rewards~\cite{chatfuzz}.
GoldenFuzz builds programs from instruction blocks, refining validity on an ISA model before using DUT coverage~\cite{goldenfuzz}.
Because such fine-tuning can overfit toward rewarded patterns, GenHuzz resets its model when coverage stalls and GoldenFuzz keeps older blocks in a recency-weighted memory~\cite{genhuzz,goldenfuzz}.
These models are learned program grammars, not mechanisms for deriving tests from a threat model.

\subsubsection{Solving from Target to Input}

Once a target net or region has been selected as described in Section~\ref{sec:target-selection}, a fuzzer can search for an activating input through mutation or solving. 
Mutation is cheap but indirect, while construction via target-to-input solving can be resource-intensive but more direct.
ProFuzz selects target nets, uses automatic test pattern generator (ATPG) to generate and merge activation patterns, and mutates the resulting seeds~\cite{profuzz}.

Software fuzzers also supplement fast mutation with specialized mechanisms for hard-to-reach paths.
Developed at Microsoft Research, SAGE achieved industry-scale symbolic execution by controlling costs through techniques such as constraint slicing and caching~\cite{sw_auto_whitebox,sw_whitebox_production}.
AFL++ includes Redqueen for simpler input-to-state correspondences~\cite{sw_redqueen,aflplusplus}.
Because RTL simulation is slow, selective solver assistance may justify its overhead, but evaluations should report solving and execution costs separately.

FormalFuzzer divides the work differently between the tool and the engineer.
Formal verification produces assertion traces, but an engineer interprets them using the ISA and design specification and chooses which fields may be mutated~\cite{formalfuzzer}.
Thus, its formal stage restricts mutation rather than directly producing a fuzz test.

\subsection{Search Effort: Seeds and Scheduling}

A useful seed may already establish protocol state, firmware configuration, privilege, or a transaction sequence that mutation is unlikely to reconstruct.
Empirical evaluations show that fuzzer performance is highly sensitive to initial seeds~\cite{encarsia,fuzz_hw_like_sw,nocfuzzer,socfuzzer}.
TaintFuzzer builds a smart-seed pool by combining inferred influence on hardware outputs with proximity to vulnerability-specific behavior~\cite{taintfuzzer}.
FUZZItizer collects a set of inputs that maximize coverage at the peripheral level, translates their bus transactions into SoC programs, and uses those programs as seeds for full-SoC fuzzing~\cite{fuzzitizer}.
NoCFuzzer found mutation more effective with shorter seeds, provided they remained long enough to express the behavior under test~\cite{nocfuzzer}.

Beyond the seeds, mutation is shaped by three policies: parent selection chooses a corpus input, energy determines the number of attempts, and operator scheduling defines the applied transformations.
DirectFuzz prioritizes inputs that cover mux-select signals in the target module and assigns more energy to inputs whose coverage is structurally closer to that module.
After coverage stalls, it restores the default energy of a low-energy input to escape a local minimum~\cite{directfuzz}.
SoCFuzzer and TaintFuzzer instead use a security-oriented cost to change mutation strategies; TaintFuzzer exhausts deterministic operators before switching to havoc~\cite{socfuzzer,taintfuzzer}.
\begin{takeaway}
The grammar and harness limit expressibility.
Seeds, construction, solving, and scheduling affect practical reachability.
Evaluations should ablate these contributions and present their verification artifacts, cost, and human effort.
\end{takeaway}

%% file: benchmarks.tex
\section{Evaluation Practices}
\label{sec:evaluation}

\begin{table*}[t]
\centering
\footnotesize
\setlength{\tabcolsep}{3pt}
\renewcommand{\arraystretch}{0.95}
\caption{Evaluation methods of 52 hardware fuzzers grouped by target design, reported metrics, and comparison with CRV.}
\label{tab:evaluation-summary}
\begin{tabular}{@{}>{\raggedright\arraybackslash}p{0.059\textwidth}p{0.028\textwidth}p{0.89\textwidth}@{}}
\hline
\textbf{Practice} & \textbf{Type} & \textbf{Details} \\
\hline
\multirow{3}{0.059\textwidth}{\raggedright Popular DUTs}
& SoC & Ariane/CVA6~\cite{socfuzzer,taintfuzzer,formalfuzzer,fuss,symbfuzz}; OpenPiton~\cite{fuzzitizer,interconfuzz,nocfuzzer}; OpenTitan~\cite{symbfuzz,interconfuzz}; Caliptra~\cite{fuzzitizer}; PicoSoC, UeRVSoC, VeeRwolf~\cite{fuss}; Custom~\cite{hyperfuzzing}. \\
& CPU & Rocket or BOOM in all but RISCover~\cite{riscover}; CVA6 follows~\cite{thehuzz,hypfuzz,psofuzz,mabfuzz,morfuzz,rlfuzz,cascade,genhuzz,goldenfuzz,refuzz,divefuzz,turbofuzz,whisperfuzz,milesan}; mor1kx/or1200~\cite{difuzzrtl,thehuzz,hypfuzz}; additional core families~\cite{processorfuzz,surgefuzz,pathfuzz,dejavuzz,divefuzz,sonar,milesan,simfuzz,refuzz,specdoctor,turbofuzz}; commercial RTL~\cite{goldenfuzz}; closed-source silicon~\cite{riscover}. \\
& IP & SiFive TileLink peripherals, FFT~\cite{rfuzz,directfuzz}; OpenTitan AES, HMAC, KMAC, RV-Timer~\cite{fuzz_hw_like_sw,fuzzwiz}; \texttt{spinal.lib}~\cite{spinalfuzz}; OpenCores and Trust-Hub IPs~\cite{vgf,profuzz,targetfuzz,synfuzz}; small CPU cores~\cite{rfuzz,directfuzz,fmtc,profuzz,targetfuzz,synfuzz}; BOOM submodules~\cite{bugsbunny}; generated FSM locks~\cite{fuzz_hw_like_sw}; netlist analysis/fuzzing~\cite{profuzz,targetfuzz,synfuzz}. \\
\hline
\multirow{3}{0.059\textwidth}{\raggedright Report Metrics}
& SoC & \textbf{Artifact: }Public infrastructure~\cite{hyperfuzzing,fuss}; repository not reproducing the evaluated stack~\cite{symbfuzz}; none found~\cite{socfuzzer,taintfuzzer,formalfuzzer,fuzzitizer,interconfuzz,nocfuzzer}.
\textbf{Coverage:} adversary-action bigrams~\cite{hyperfuzzing}; branch and toggle~\cite{fuss}; VCS code coverage~\cite{fuzzitizer}; control-register state~\cite{symbfuzz,interconfuzz}; line, branch, UVM covergroups~\cite{nocfuzzer}; none, cost-function convergence~\cite{socfuzzer,taintfuzzer,formalfuzzer}. \textbf{Progress:} executions~\cite{socfuzzer,taintfuzzer,formalfuzzer}, vectors~\cite{symbfuzz}, packets~\cite{interconfuzz}, programs~\cite{fuzzitizer}, or wall-clock~\cite{fuss,nocfuzzer}.
\textbf{Comparison with other fuzzers:} Self-ablation~\cite{hyperfuzzing,socfuzzer,taintfuzzer,formalfuzzer,interconfuzz,nocfuzzer,fuss}; same-lineage rerun~\cite{formalfuzzer}; artifact rerun~\cite{symbfuzz}; reimplementation under one coverage tool~\cite{fuzzitizer}; numbers from its paper~\cite{symbfuzz,interconfuzz}; feature table~\cite{socfuzzer,taintfuzzer,formalfuzzer}; no compatible prior fuzzer~\cite{nocfuzzer}.
\\
& CPU & \textbf{Artifact:} Public code~\cite{difuzzrtl,presifuzz,morfuzz,surgefuzz,pathfuzz,cascade,divefuzz,sigfuzz,dejavuzz,phantomtrails,milesan,riscover,processorfuzz,thehuzz,hypfuzz,psofuzz,mabfuzz,refuzz,rlfuzz,specdoctor}; partial/placeholder~\cite{simfuzz,genfuzz,genhuzz,whisperfuzz}; promised, no URL~\cite{turbofuzz}; on request~\cite{goldenfuzz}; none found~\cite{chatfuzz,introspectre,specure,sonar,geier2025}.
\textbf{Coverage:} control-register state~\cite{difuzzrtl,morfuzz,genfuzz,turbofuzz,cascade,divefuzz,simfuzz}; ISS CSR transitions~\cite{processorfuzz}; commercial-simulator metrics~\cite{thehuzz}, narrowed to a subset by successors~\cite{hypfuzz,psofuzz,mabfuzz,refuzz,chatfuzz,genhuzz,goldenfuzz,rlfuzz}; taint, leakage, timing, or contention signals~\cite{dejavuzz,phantomtrails,specure,whisperfuzz,sonar,geier2025}; none~\cite{specdoctor,milesan,riscover}. \textbf{Progress:} wall-clock or test count.
\textbf{Comparison with other fuzzers:} Artifact-derived~\cite{processorfuzz,divefuzz,simfuzz}; baseline tests replayed in-house~\cite{morfuzz,dejavuzz}; reimplementation~\cite{difuzzrtl,psofuzz,mabfuzz,surgefuzz,cascade}; numbers taken from prior work~\cite{mabfuzz,rlfuzz,pathfuzz,phantomtrails,milesan,cascade}; baseline embedded as component~\cite{hypfuzz,refuzz}; provenance unstated~\cite{genfuzz,chatfuzz,genhuzz,goldenfuzz,turbofuzz,sonar,specure}; feature table~\cite{specdoctor,sigfuzz,whisperfuzz}.
\\
& IP & \textbf{Artifact:} Public code~\cite{rfuzz,rtlfuzzlab,spinalfuzz,fuzz_hw_like_sw}; promised, no URL~\cite{directfuzz}; none found~\cite{fmtc,profuzz,targetfuzz,synfuzz,fuzzwiz,bugsbunny,vgf}; commercial ATPG/synthesis/simulation licenses also needed~\cite{profuzz,targetfuzz,synfuzz}.
\textbf{Coverage:} mux-select toggle~\cite{rfuzz,directfuzz}, stricter under the same name~\cite{fmtc}; Verilator line or basic block~\cite{spinalfuzz,fuzzwiz,fuzz_hw_like_sw}; netlist library-cell toggle~\cite{synfuzz}; target activation~\cite{profuzz,targetfuzz}; bespoke value, distance, or flow-path metrics~\cite{vgf,bugsbunny}. Ten report no bugs; known benchmark faults~\cite{vgf}; new synthesis bugs and CVEs~\cite{synfuzz}.
\textbf{Comparison with other fuzzers: } Fork of the baseline codebase~\cite{directfuzz}; reimplementation on another toolchain~\cite{fmtc,targetfuzz}; external fuzzer on shared designs~\cite{vgf}; numbers from its paper~\cite{profuzz,fuzzwiz}; self-ablation only~\cite{rfuzz,bugsbunny}; commercial equivalence checker~\cite{synfuzz}; feature table~\cite{synfuzz,fuzzwiz}; no evaluation~\cite{rtlfuzzlab}. \\
\hline
\multirow{3}{0.059\textwidth}{\raggedright Compared to CRV}
& SoC & Explicit CRV in the same UVM environment~\cite{nocfuzzer}; UVM random testing~\cite{symbfuzz}; uniform random regression~\cite{fuzzitizer}; random without symbolic execution~\cite{interconfuzz}; CRV as motivation only~\cite{fuss}; none~\cite{hyperfuzzing,socfuzzer,taintfuzzer,formalfuzzer}. \\
& CPU & riscv-dv, on one core only~\cite{morfuzz}; riscv-torture~\cite{difuzzrtl,morfuzz,surgefuzz}; random regression~\cite{thehuzz,hypfuzz}; generators used only as a seed corpus~\cite{pathfuzz,simfuzz}; guidance-disabled ablation, not CRV~\cite{processorfuzz,genfuzz,refuzz,turbofuzz,introspectre,sonar,geier2025,riscover}. \\
& IP & CRV in the same testbench~\cite{spinalfuzz}; CRV on generated FSM locks only~\cite{fuzz_hw_like_sw}; regression speedup cited, not run~\cite{fuzzwiz}; uniform random~\cite{rfuzz}; random only as seeds~\cite{fmtc,bugsbunny}; none~\cite{directfuzz,profuzz,targetfuzz,synfuzz,rtlfuzzlab,vgf}. \\
\hline
\end{tabular}
\end{table*}

Hardware-fuzzer evaluations commonly report bugs, time to detection, and achieved coverage.
However, the evidence is difficult to compare across fuzzers~\cite{encarsia}, and rarely establishes whether fuzzing improves upon modern CRV.
Our framework provides a systematic basis for determining when such comparisons are meaningful.
Evaluations should first align the verification objective, target abstraction, oracle, and budget, and then isolate the effects of guidance and input generation.
Table~\ref{tab:evaluation-summary} summarizes current evaluation practice.

\subsection{Comparability between Fuzzers}
\label{sec:compare-fuzzers}

Direct comparison requires the same verification objective and target abstraction.
A CPU campaign for transient-execution leakage could be compared with campaigns for the same leakage model, but not with IP-level functional coverage closure.
The DUT name alone is also insufficient because its configuration and revision determine its behavior and bug set.
Many studies name Rocket, BOOM, or OpenTitan without identifying the evaluated revision.
Encarsia likewise identifies design and version diversity as a barrier to comparing naturally occurring bug counts~\cite{encarsia}.

Aligned verification objectives can still use different oracles.
SpecDoctor combines a rollback monitor with differential timing tests, DejaVuzz uses differential information-flow tracking and taint-liveness annotations, and MileSan compares architectural and microarchitectural information flows and requires leakage to affect instruction timing~\cite{specdoctor,dejavuzz,milesan}.
These checks recognize different failure sets despite their shared high-level goal.
Evaluations of guidance or input generation should therefore use the same oracle whenever possible.
Otherwise, the result compares complete campaigns and cannot attribute an advantage to the search mechanism alone.
The oracle specification, observations, and blind spots must then be reported as experimental variables.

Once these conditions align, component ablations should change one choice at a time.
Guidance includes observations, feedback, retention, and scheduling.
Input generation includes the harness, grammar, corpus, mutation or construction, and solver assistance.
Encarsia demonstrates their interaction.
Two CPU fuzzers found the same injected bugs with structural feedback enabled and disabled, while different seed programs changed which bugs were found~\cite{encarsia}.

Coverage does not ensure a bug-free design, but maximizing a widely accepted structural or functional metric could be a meaningful contribution when coverage closure is the stated objective.
Such work should compare fuzzers with the same objective and abstraction, starting from a common corpus and evaluating their generated tests with the same coverage tool and configuration.
Replaying fuzzer-generated inputs into the DUT under the independent metric separates evaluation coverage from each fuzzer's internal feedback.

Initial corpora and pseudorandom seeds require explicit control.
Compared fuzzers should start from the same corpus when their grammars permit it.
Studies should report distributions and uncertainty over multiple independent seeds because fuzzing performance is seed-sensitive~\cite{encarsia,fuzz_hw_like_sw,nocfuzzer}.
A reproducible artifact should preserve these campaign-level control variables and the execution scripts.
If artifacts cannot be released, papers should compensate with comprehensive reporting of coverage metrics, effectiveness results, execution budgets, and compute and resource costs. 
If the fuzzer itself cannot be released, the complete, ordered set of inputs generated over each evaluated campaign should also be released so others can replay the baseline, measure coverage independently, and reproduce bug findings.
Sec.~\ref{sec:user-perspective} discusses the deployment requirements.

\subsection{Comparison with CRV}
\label{sec:compare-crv}

Few hardware fuzzers compare against modern CRV.
When doing so, uniform random generation and disabling guidance are useful ablations. 
However, neither is an adequate baseline for modern CRV, which combines constrained stimulus generation with a verification plan, functional coverage model, and oracle.
Fuzzers that seek broad functional bugs or coverage closure are proposing to automate part of this CRV workflow and should compare against it directly~\cite{rfuzz,difuzzrtl,processorfuzz,spinalfuzz,nocfuzzer}.

Comparisons should give each approach the same time and compute budget, then measure the coverage achieved and bugs found rather than the number of tests executed.
Fuzzing can incur greater per-test overhead than CRV due to guidance and input generation overhead.
But fuzzing could be better than CRV if fewer tests close important coverage targets or expose failures.
Results should state the targets and conditions under which either approach wins rather than give a universal ranking.
NoCFuzzer provides a rare example by comparing fuzzing and CRV in the same UVM environment using shared coverage targets~\cite{nocfuzzer}.

We propose a comparison based on the target design's existing regression when one is available.
Run the nightly CRV regression and record its duration, compute resources, simulator configuration, oracle outcomes, and engineer-defined functional coverage.
Give the fuzzer the same time and compute budget, record the inputs it generates, and replay them through the CRV environment with the functional covergroups and oracle illustrated in Algorithm~\ref{alg:crv}.
This replay measures test quality with the same engineer-defined functional coverage model and oracle instead of the fuzzer's internal coverage.
The comparison should also report memory use and the manual effort for CRV constraints and for fuzzer-specific harnesses, seeds, properties, and instrumentation.

Known-bug experiments provide a complementary result.
The fuzzer and CRV regression should run against the same faulty revision or injected bugs and report the number found and detection-time distributions.
Reporting whether each bug was reached, triggered, and detected helps distinguish failures of input generation from blind spots in the oracle.
Similarly, directed adversarial fuzzers should be compared against baselines targeting the same security objective.
A generic coverage regression cannot establish whether a fuzzer is better at triggering a transient-execution leak or violating an information-flow property unless the CRV campaign is configured to seek that same behavior.

\subsection{Proposed Evaluation Practices}
\label{sec:proposed-evaluation}

Every evaluation should identify the DUT commit and configuration, verification objective, target abstraction, oracle, corpus, time and resource budget, tools, and measurement conditions.
It should also preserve raw results and report distributions from repeated trials.
Coverage, time to failure, resource use, and user effort measure different aspects of a campaign and should be reported separately.
Claims of generality require varied representative objectives and target abstractions.

Software fuzzing faced the same fragmented evaluation landscape.
Klees et al. found problems in every one of 32 surveyed evaluations and showed that trials, seeds, timeouts, targets, statistics, and metrics can change conclusions~\cite{sw_evaluating}.
FuzzBench turned these lessons into a service with versioned real targets, common corpora, equal resources, repeated trials, independent coverage, raw data, and statistical reports~\cite{sw_fuzzbench}.
Magma added bug ground truth by reintroducing real bugs into real programs and distinguishing reach, trigger, and detection~\cite{sw_magma}.

Adapting the software benchmark model to hardware requires benchmark families organized by target abstraction and verification objective rather than a single leaderboard.
A standardized benchmark should package versioned DUT configuration with its harness, corpus, oracle, coverage models, faulty and fixed revisions, injected bugs, and replay scripts.
It should define resource limits and measurement procedures while allowing each fuzzer to retain its own internal guidance.
Separate tracks could cover a suite of security IPs, an industry-scale NoC, a complex CPU, and a full SoC.
OpenTitan, Caliptra, OpenPiton, and BOOM are promising starting points because they span these abstractions and are already used by multiple surveyed fuzzers~\cite{difuzzrtl,nocfuzzer,fuzzitizer,symbfuzz}.
OpenTitan also provides extensive open verification assets~\cite{opentitan_dv_methodology}.

The bug set should combine documented real bugs with reproducible injected bugs.
Hack@DAC designs already provide competition-injected security bugs in PULPissimo, OpenPiton, and OpenTitan, while EnCorpus provides 90 formally checked CPU bugs~\cite{assertion_benchmarks,symbfuzz,encarsia}.
Existing SystemVerilog assertion suites can serve as shared oracles for some of these benchmark bugs~\cite{assertion_benchmarks}. 
Providing both faulty and fixed DUT revisions allows researchers to confirm that a reported failure is caused by the benchmarked bug.
Using our framework, benchmarks should label each bug by its verification objective, target abstraction, and oracle, allowing readers to distinguish out-of-scope bugs from in-scope bugs not reached or detected.

Automatic property generation could eventually provide a common set of properties and violations, but it remains an open problem~\cite{hwsec_properties}. 
For now, the practical alternative is a versioned benchmark with curated assertions, reference models, covergroups, and known bugs.
Building them will require substantial community effort and no initial suite will satisfy every objective.
A shared benchmark service would nevertheless be a major contribution and help establish the evidence and infrastructure needed for the widespread adoption of hardware fuzzing.

\begin{takeaway}
Fair evaluation aligns the objective, target abstraction, DUT version, oracle, and budget before comparing search mechanisms.
Coverage-maximizing fuzzers should be measured against established CRV, while directed campaigns need baselines with the same security objective.
The community should build shared, versioned benchmarks to make these comparisons reproducible.
\end{takeaway}

%% file: how_to_use.tex
\section{Deployment and Reproducibility}
\label{sec:user-perspective}

The reproducibility challenges discussed in Section~\ref{sec:evaluation} stem partly from deployment requirements that extend beyond the fuzzer itself. 
Users may need to supply or reconstruct target-specific artifacts, an oracle, toolchains, and access to the design or execution platform, affecting both usability and reproducibility. 
We examine these requirements as the artifacts needed to construct inputs and specify verification objectives, and the infrastructure access to run campaigns.

\subsection{Required Artifacts}

The initial deployment hurdle involves providing the design artifacts and input-generation artifacts required by the fuzzer. At the structural level, nearly all frameworks require access to a representation of the target hardware, but the required representation varies across methodologies. Some operate on high-level hardware descriptions such as Chisel, Flexible Intermediate Representation for RTL (FIRRTL), and SpinalHDL~\cite{spinalfuzz,specdoctor,difuzzrtl}, while others require conventional Verilog or SystemVerilog RTL~\cite{formalfuzzer,symbfuzz,hypfuzz}. More specialized approaches require additional artifacts, such as synthesized netlists and standard-cell libraries~\cite{synfuzz}, ATPG-specific configuration~\cite{profuzz}, or complete UVM environments for NoC verification~\cite{interconfuzz,nocfuzzer}.

The harness and seed corpus artifacts required to generate valid inputs likewise vary with the target abstraction. 
Processor fuzzers rely on program-aware representations~\cite{difuzzrtl,thehuzz,morfuzz,divefuzz}, whereas SoC and interconnect fuzzers~\cite{socfuzzer,taintfuzzer,formalfuzzer,interconfuzz,nocfuzzer} and RTL-IP fuzzers~\cite{fuzzwiz,fuzz_hw_like_sw,targetfuzz} depend on target-specific input models and harnesses. 
As Section~\ref{sec:input-generation} details, the performance of fuzzers and the failures they can find within a limited budget are highly sensitive to these artifacts, making them a critical part of deployment.

CRV-style verification can build on established verification flows and reusable infrastructure, whereas directed adversarial testing typically requires more manual effort to translate campaign-specific threat models and security specifications into properties that various tools can use.
Formal- and property-driven approaches require users to encode behaviors or properties that guide exploration~\cite{hyperfuzzing,symbfuzz,formalfuzzer,hypfuzz}. 
Microarchitectural security fuzzers may require threat models, speculation gadgets, taint annotations, or contention templates~\cite{specdoctor,introspectre,specure,dejavuzz,milesan,sonar}. 
In these settings, encoding the security assumptions and specifications that define the search space accounts for most of the user effort.

Recent machine-learning and optimization-based approaches extend this requirement to substantial data preparation and configuration. 
Large language model (LLM)-based frameworks may require large instruction datasets, while reinforcement-learning and optimization-based approaches rely on auxiliary state representations, reward formulations, seed-generation configurations, optimization parameters, or historical test suites~\cite{chatfuzz,genhuzz,goldenfuzz,rlfuzz,mabfuzz,psofuzz,refuzz}. 
Consequently, the artifacts required to deploy a fuzzer can extend well beyond the DUT and a seed corpus, with user effort increasingly determined by the domain-specific information required by the methodology.

\begin{table}[hb]
\centering
\caption{Visibility models adopted across the 52 fuzzers.}
\label{tab:visibility}
\footnotesize
\include{tables/how_to_use/visibility}
\end{table}

\subsection{Infrastructure and Design Visibility}

A primary challenge in deploying hardware fuzzers is the toolchain infrastructure and specialized hardware required to execute them. 
A substantial portion of the analyzed CPU, SoC, and NoC fuzzers depend on proprietary simulation or formal verification tools, most notably Synopsys VCS and Cadence JasperGold/Xcelium~\cite{fuzzitizer,nocfuzzer,profuzz,synfuzz,whisperfuzz,specure,presifuzz,morfuzz,hypfuzz,thehuzz,mabfuzz,chatfuzz,genhuzz,goldenfuzz,refuzz,rlfuzz}. 
Open-source simulators such as Verilator are widely used for RTL-IP modules and lightweight CPU targets~\cite{hyperfuzzing,fuss,rfuzz,directfuzz,fuzzwiz,bugsbunny,spinalfuzz,difuzzrtl,processorfuzz,cascade,divefuzz}, but they often require additional co-simulation infrastructure, including custom C++ harnesses and external libraries such as Z3, PyVerilog, or \texttt{cocotb}~\cite{formalfuzzer,fuss,symbfuzz,interconfuzz,rtlfuzzlab,fmtc,geier2025,specure,difuzzrtl}. 
Most academic fuzzers also require users to adapt their designs, testbenches, and verification assets to a custom fuzzing environment rather than integrating directly into an existing industry-standard workflow such as a UVM-based environment~\cite{hwfuzzenv}.

These requirements can extend beyond the simulation environment itself. 
Frameworks that rely on physical or hardware-accelerated execution may require specialized hardware platforms, while machine learning (ML)- and LLM-guided approaches can introduce substantial compute and software dependencies in addition to conventional UVM infrastructure~\cite{socfuzzer,taintfuzzer,turbofuzz,riscover,chatfuzz,genhuzz,goldenfuzz,rlfuzz}. 
Thus, deployment may require not only software toolchains but also dedicated hardware and compute environments that are not part of the fuzzer.

These infrastructure requirements are closely coupled with the level of internal design access required by the fuzzer. 
As shown in Table~\ref{tab:visibility}, most surveyed frameworks use grey-box or white-box visibility. 
Grey-box approaches use lightweight instrumentation to obtain runtime feedback, while white-box approaches require access to internal design artifacts such as RTL, netlists, or data flow and may require users to integrate assertions, sanitizers, or information-flow tracking logic. 
Black-box approaches avoid internal instrumentation but provide less internal observability and may rely on physical execution environments. 
Thus, the visibility model determines both what information the fuzzer can exploit and what design access and instrumentation the user must provide.

From a reproducibility perspective, these requirements mean that reproducing a published campaign may require substantially more than the fuzzer implementation itself. 
Researchers may also need the exact DUT and configuration, input artifacts, tool versions or licenses, supporting verification infrastructure, and specialized hardware. 
When these components are unavailable or unreleased, independent reproduction becomes difficult even when the design is public.

\begin{takeaway}
Hardware fuzzing should move towards well-defined, stable interfaces between reusable fuzzer engines and target-specific verification assets.
Researchers should release these assets to improve reproducibility, support fair comparisons, and accelerate industry adoption.
\end{takeaway}

%% file: tables/how_to_use/visibility.tex
\begin{tabular}{lp{0.68\columnwidth}}
\toprule
\textbf{Visibility} & \textbf{Works} \\
\midrule

Black-box (2) &
\cite{riscover,simfuzz} \\

\multirow{2}{*}{Grey-box (30)}
& \cite{socfuzzer,taintfuzzer,formalfuzzer,rfuzz,directfuzz,
profuzz,targetfuzz,synfuzz,rtlfuzzlab,fmtc,genfuzz,morfuzz,hypfuzz,thehuzz,psofuzz} \\
& \cite{fuzz_hw_like_sw,fuzzwiz,bugsbunny,spinalfuzz,vgf,sigfuzz,
geier2025,difuzzrtl,presifuzz,processorfuzz,mabfuzz,
chatfuzz,pathfuzz,cascade,divefuzz} \\
\multirow{2}{*}{White-box (20)}
& \cite{hyperfuzzing,fuss,fuzzitizer,symbfuzz,
interconfuzz,nocfuzzer,specdoctor,turbofuzz,goldenfuzz,refuzz,rlfuzz} \\
& \cite{whisperfuzz,dejavuzz,phantomtrails,milesan,
introspectre,specure,sonar,surgefuzz,genhuzz} \\

\bottomrule
\end{tabular}

%% file: discussion.tex
\section{Summary and Conclusion}

In this paper, we introduce a framework developed through an analysis of 52 hardware fuzzers.
The framework treats verification as a bounded search shaped by the verification objective, campaign budget, target abstraction, and the components of the fuzzer.
We then examine how software fuzzing grew from a novel testing technique into a mature research field with widespread industry adoption.
This history provides lessons that can help advance hardware fuzzing.

Our central takeaway is \emph{you find what you seek}.
A fuzzing campaign finds failures only where its oracle, input generator, and guidance overlap and align with its verification objective, within its limited budget.
The oracle defines the set of failures that can be detected, guidance prioritizes tests for further exploration, and input generation produces new tests.
All three mechanisms individually and collectively define what the campaign seeks, and improving one component cannot remove the limits imposed by the others.
Generic structural feedback reduces manual effort, but it is meaningful only when it preserves behavior relevant to the objective.
Coverage closure is informative when the covered items correspond to a verification plan, while directed progress requires a target and score derived from a property or threat model.
Quality guidance therefore depends on preserving relevant distinctions from the verification objective through target behavior, observation, feedback, and search policy.
Hardware fuzzers should therefore strive for strong alignment among their oracle, guidance, and input generation so that every component directs the campaign toward the intended failures.
Researchers should also perform subcomponent-level ablations to understand how each choice affects the fuzzer's behavior.

The same principle applies to target-specific verification assets, which define what a campaign can generate, reach, and observe.
Fuzzers, like other search and optimization engines, are likely to be better suited to particular applications and bug classes.
The community should establish a comprehensive family of benchmarks across the IP, CPU, NoC, and SoC designs to identify these relative strengths.
Each benchmark should include versioned DUTs and the complete verification and fuzzing collateral needed to reproduce a campaign, including seeds, covergroups, oracles, injected and real bugs, and replay scripts.
It should also define target metrics, resource budgets, and testing procedures for fair comparisons.
Coverage-maximizing fuzzers should be compared with established CRV in the same environment, with the manual effort for both approaches reported.
Directed adversarial campaigns should show that their guidance and input generation reach a specified violation more effectively than alternatives with a comparable objective, threat model and budget.

Finally, industry adoption requires reusable tools that support continuous fuzzing, failure triage, and reliable replay across platforms.
Software fuzzing became sustainable through reusable engines, stable interfaces, documentation, standardization, open-source development practices, continuous triage, and shared evaluation infrastructure~\cite{aflplusplus,oss_fuzz,sw_fuzzbench}.
Hardware verification already maintains target-specific UVM sequences, constrained generators, assertions, scoreboards, reference models, functional covergroups, and threat models.
Researchers who seek adoption should make their tools easy to integrate with these workflows and release the assets needed to reproduce and extend their work.
As a community, we should address the unique challenges of hardware verification by drawing lessons from the path established by software fuzzing.
For hardware fuzzing to become mainstream, its components must align with the verification objective, integrate with existing verification assets through stable interfaces, and be evaluated through reproducible campaigns that report effectiveness, cost, and required user effort.